\documentclass[aps,prb,twocolumn,showpacs]{revtex4}
\usepackage{graphicx}

\begin{document}

\title{Electronic structures and ferromagnetism in transition metals codoped ZnO}

\author {Min Sik Park and B. I. Min}

\affiliation{Department of Physics and electron Spin Science Center,  \\
Pohang University of Science and Technology, Pohang 790-784, Korea}

\date{\today}

\begin{abstract}

We have investigated electronic structures and magnetic properties of
potential ZnO based diluted magnetic semiconductors: 
(Fe, Co) and (Fe, Cu) codoped ZnO.
The origins of ferromagnetism are shown to be
different between two. (Fe, Co) codoped ZnO does not have a tendency 
of Fe-O-Co ferromagnetic cluster formation,
and so the double exchange mechanism will not be effective.
In contrast, (Fe, Cu) codoped ZnO has a tendency of
the Fe-O-Cu ferromagnetic cluster formation with the
charge transfer between Fe and Cu, which would lead to the ferromagnetism 
through the double-exchange mechanism.
The ferromagnetic and nearly half-metallic ground state is obtained
for (Fe, Cu) codoped ZnO.

\end{abstract} 


\maketitle


\section{Introduction}
\label{sec:intro}
   
Electronics utilizing the spin degree of freedom of electrons, namely
``spintronics'',
becomes an important emerging field. The spintronics is
expected to overcome the limits of traditional microelectronics, 
such as non-volatility, data processing speed, electric power consumption, 
and integration densities\cite{Wolf}.
Diluted magnetic semiconductors (DMSs) will play a core role in 
spintronics as semiconductors do in electronics
due to easy integration into existing electronic devices.
Two types of DMS families have been well studied:
II-VI type such as Mn-doped CdTe and ZnSe\cite{Furdyna}, 
and III-V type such as Mn-doped GaAs\cite{Ohno}. 
In particular, 
the latter attracts great attention, because it becomes a ferromagnetic (FM) 
DMS having the Curie temperature $T_{C} \sim 110K$. 
Recent research effort has been 
focused on developing a new FM-DMS operating at
room temperature\cite{Dietl,Sato,Medvedkin,Matsumoto,Ando,SCho}. 

Along this line, attempts have been made to fabricate ZnO based DMS.
ZnO is a wide-gap (E$_g$ $\sim$ 3.44eV) II-VI semiconductor, and
so it can be used for ultraviolet light emitting devices. 
Jin {\it et al.}\cite{Jin} fabricated 3$d$ transition metal (TM) doped 
epitaxial ZnO thin films using the combinatorial laser molecular-beam epitaxy 
method. However, they have not detected any indication of ferromagnetism. 
In contrast, Ueda {\it et al.}\cite{Ueda} observed the FM behaviors
in some of the Co-doped ZnO films made by using the pulsed-laser
deposition technique with T$_C$ higher than the room temperature. 
The reproducibility, however, was less than 10$\%$.
Hence the realization of the FM long range order in
Co-doped ZnO films is controversial.

On the other hand,
there were also trials to make ZnO based DMS by co-doping two TM elements:
(Fe,Co) or (Fe,Cu).
$\rm Zn_{1-x}(Co_{0.5}Fe_{0.5})_xO$ films were fabricated 
by using the reactive magnetron co-sputtering technique\cite{Cho},
which seemed to have the single phase of the same wurtzite 
structure as pure ZnO up to $x$=0.15.
The room temperature ferromagnetism was observed and
the rapid thermal annealing under vacuum 
leads to increases in $T_{C}$, magnetization, and the carrier concentration.
Also Cu-doped $\rm Zn_{1-x}Fe_{x}O$ bulk samples were fabricated\cite{Han}. 
The bulk sample has the advantage of insensitivity to the detailed process 
conditions, over the film sample fabricated under the nonequilibrium 
condition.  They observed the FM behaviors with
T$_C$$\sim$550K and the saturation magnetic moment
of 0.75$\mu_B$ per Fe in $\rm Zn_{0.94}Fe_{0.05}Cu_{0.01}O$. 
The saturation magnetic moment increases,
as the Cu doping ratio increases up to 1$\%$.
In addition, the large magnetoresistance was observed below 100K.

Motivated by these reports of ferromagnetism in TM codoped ZnO,
we have studied electronic structures and magnetic properties of 
(Fe, Co) and (Fe, Cu) codoped ZnO:
$\rm Zn_{0.875}(Fe_{0.5}{\it M}_{0.5})_{0.125}O$ ({\it M}=Co or Cu).
We have used the linearized muffin-tin orbital (LMTO) band method in the 
local spin-density approximation (LSDA). 
To explore the Coulomb correlation and the spin-orbit (SO) effect 
in TM-doped ZnO, we have also employed the LSDA+$U$(+SO) method
incorporating the Coulomb correlation
interaction $U$ and the SO interaction\cite{Kwon}.
ZnO has the wurtzite structure in which anions and cations form hexagonal
close-packed lattices. The wurtzite ZnO is composed of
tetrahedrons formed by four O anions.
For $\rm Zn_{0.875}(Fe_{0.5}{\it M}_{0.5})_{0.125}O$, we have considered
an orthorhombic supercell containing sixteen formula units in the
primitive unit cell by replacing two Zn atoms by Fe and {\it M} atoms 
($\rm Zn_{14}Fe_1{\it M}_1O_{16}$). For the lattice constants,
we assumed those of pure ZnO with $a=6.4998, b=11.2580, c=5.2066$ $\rm \AA$.

\begin{figure}[t]
\includegraphics[scale = 0.4]{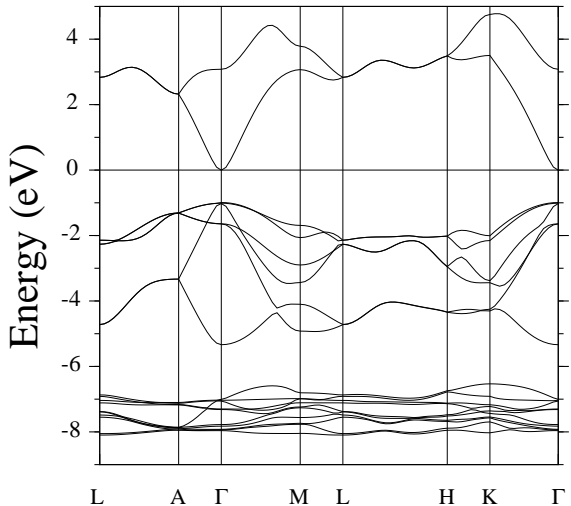}
\caption{The LSDA+$U$ band structure of ZnO with $U$=3.0eV
for Zn $3d$ electrons. The lowest and intermediate 
bands correspond to mainly Zn $3d$ and O $2p$ bands, respectively.
Zn $3d$ bands in the LSDA are too shallow to become mixed with
O $2p$ bands.}
\label{zno}
\end{figure}
\begin{figure}[t]
\includegraphics[scale = 0.4]{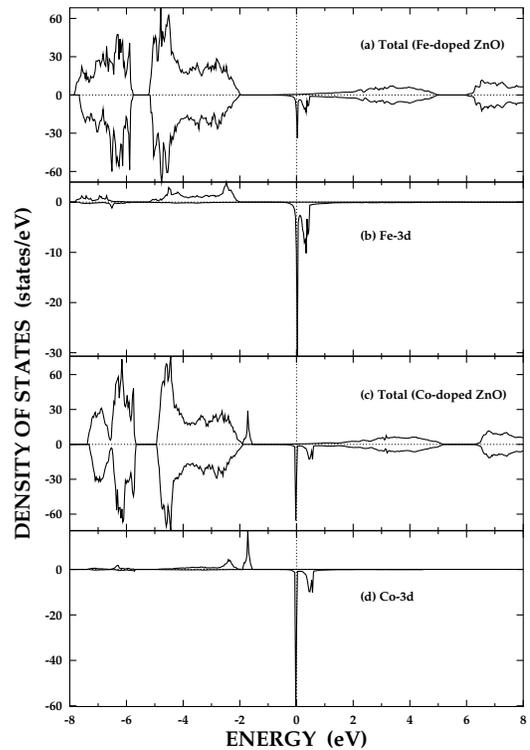}
\caption{The LSDA total and TM 3$d$ PLDOSs
         of $\rm Zn_{0.9375}Fe_{0.0625}O$ and 
            $\rm Zn_{0.9375}Co_{0.0625}O$.
}
\label{feco}
\end{figure}
\begin{figure}[t]
\includegraphics[scale = 0.4, angle=270]{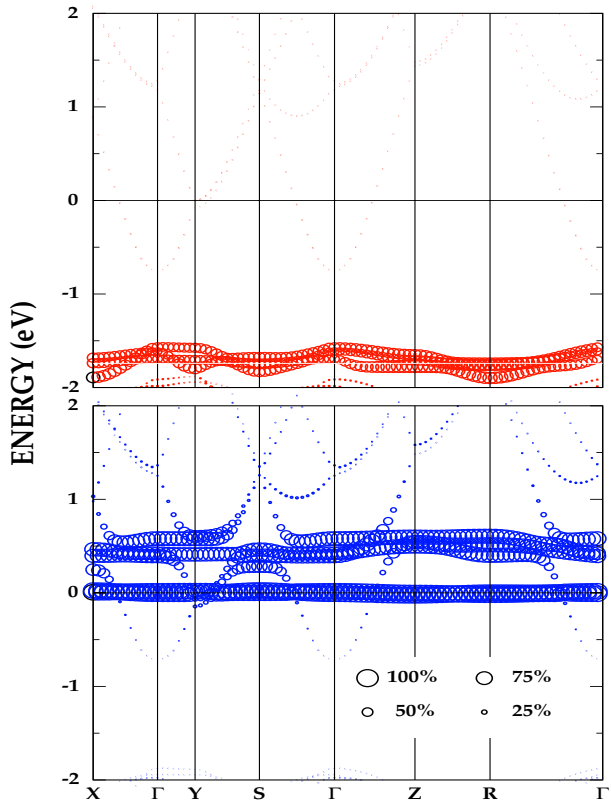}
\caption{Band structure of $\rm Zn_{0.9375}Co_{0.0625}O$ near the Fermi level
         (the upper panel for the majority and the lower for the minority
          spin band). The size of circle represents the amount of Co 3$d$
         component in the wave function. 
}
\label{band}
\end{figure}
\begin{figure}[t]
\includegraphics[scale = 0.4]{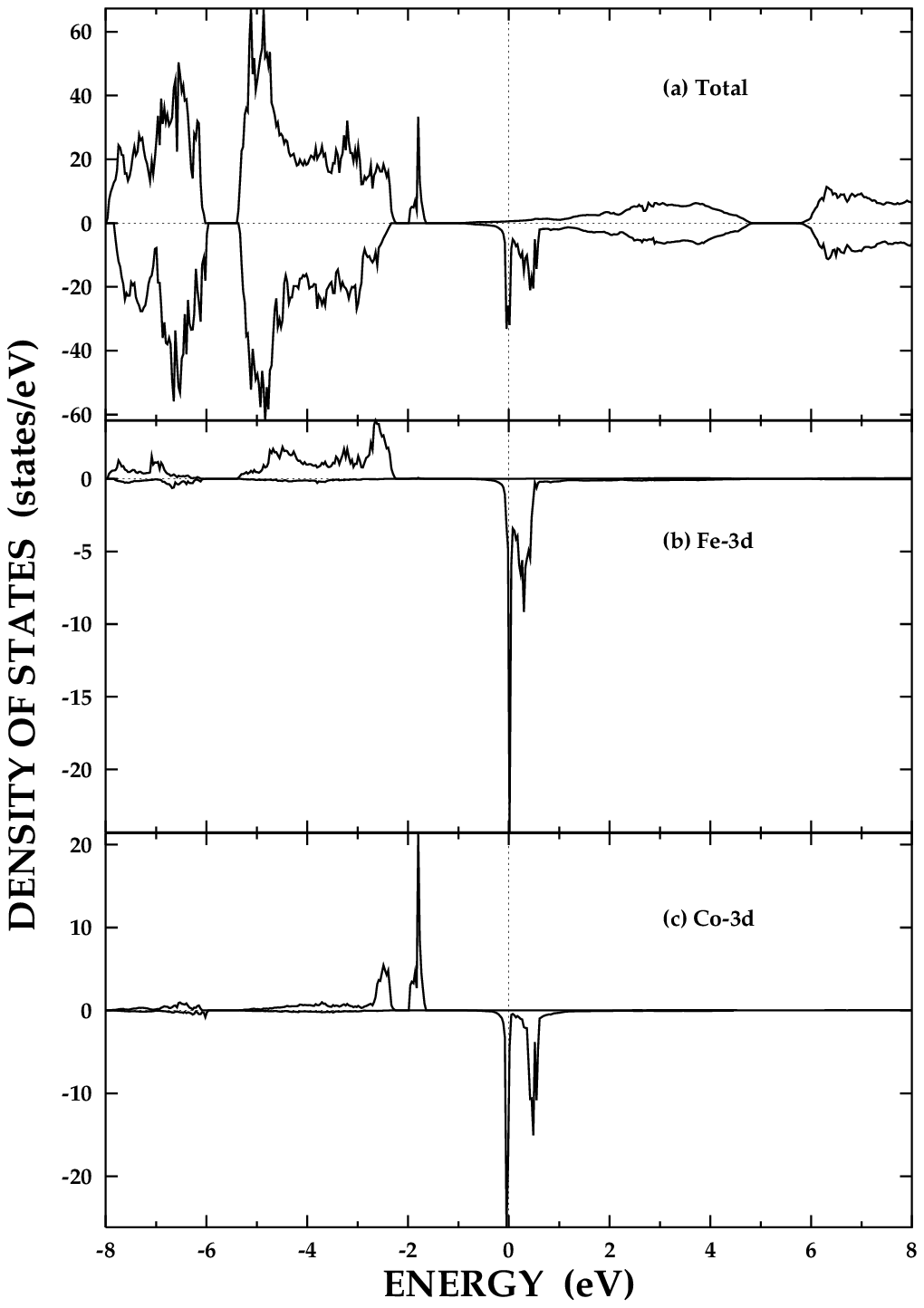}
\caption{The LSDA total and TM 3$d$ PLDOSs
        of $\rm Zn_{0.875}Fe_{0.0625}Co_{0.0625}O$.
}
\label{zfco}
\end{figure}

\section{Z\lowercase{n}O, F\lowercase{e}- \lowercase{and} 
C\lowercase{o}-\lowercase{doped} Z\lowercase{n}O}

First, we have checked the electronic structure of pure wurtzite ZnO
without doping elements. The overall band structure of
the present LMTO result is consistent with existing results
\cite{Schroer,Arya,Usuda}.
As usual in the LSDA calculations,
the obtained energy gap $\sim 0.7$ eV is only about 20$\%$ of
the experimental value.
Also the position of Zn 3$d$ band ($-3.0 \sim -5.0$ eV with
respect to the valence band top) is much shallower
than the Zn 3d spectrum obtained by photoemission experiment
($\sim -9.0$ eV) \cite{Ruckh}.
The LSDA+$U$ band method improves the results of LSDA, but not
so satisfactorily. As shown in Fig.~\ref{zno}, 
the energy gap is increased to $\sim 1.0$ eV,
and the position of Zn 3$d$ band becomes deeper ($-6.0 \sim -7.0$ eV).
The LSDA+$U$ result for Zn 3$d$ position is consistent with result 
obtained by the GW band calculation\cite{Arya,Usuda}.
Still the energy gap and Zn 3$d$ position are smaller and shallower 
than experimental values.

We then have performed the LSDA band calculations for single TM-element
doped ZnO: Fe- and Co-doped ZnO with 6.25$\%$ concentration of
TM element. Due to heavy computational load,
we have not considered the Coulomb correlation $U$ for Zn $3d$ electrons 
in the supercell calculations of doped ZnO systems\cite{Super}.
Note that electronic structures of TM doped ZnO have already been 
reported for 25$\%$ concentration of TM element
by using the Korringa-Kohn-Rostoker band method combined with the
coherent potential approximation\cite{Sato}. 
They found that V-, Cr-, Fe-, Co-, and Ni-doped ZnO
would have the FM ground states rather than the spin-glass
states. With this background, we thus consider below only the FM states
for Fe-, and  Co-doped ZnO with more realistic TM doping
concentration.

Figure~\ref{feco} shows the LSDA total density of states (DOS)
and TM 3$d$ projected local DOS (PLDOS) for
$\rm Zn_{1-x}Fe_{x}O$ and those for $\rm Zn_{1-x}Co_{x}O$ $(x=0.0625)$.
For both Fe- and Co-doped ZnO, we have obtained nearly $half$-$metallic$ 
electronic structures, that is, the conduction electrons 
at the Fermi level $E_F$ are almost 100$\%$ spin polarized.
The minority spin $3d$ states of both Fe and Co near $E_F$ are seen 
to be hybridized slightly with the conduction band (see Fig.~\ref{band}).
Both for Fe-doped and Co-doped ZnO, the Fermi level cuts the sharp 
minority spin $e_g$ states. Note that Fe and Co are located at
tetrahedral centers formed by O ions, and so $e_g$ states are lower in
energy than $t_{2g}$ states.
Since there is nearly one more $d$-electron in Co-doped ZnO,
the Fermi level is located near the valley between
Co $e_g$ and $t_{2g}$ minority spin state.
The exchange splittings are larger than the crystal field splittings, 
reflecting high spin states of Fe and Co in ZnO. 
Especially, the exchange splitting
in Fe-doped ZnO is very large to locate both $e_g$ and $t_{2g}$ states deep
in energy, and so the majority spin Fe $t_{2g}$ states become 
fully hybridized with O $2p$ states to yield a broad band. 
In contrast, the $t_{2g}$ states in Co-doped ZnO
are shallow and located above the O $2p$ valence band and so 
the hybridization becomes weak.

The total magnetic moments 4.44 and 3.25 $\mu_B$ for Fe- and Co-doped ZnO
come mostly from Fe (4.11 $\mu_B$) and Co (2.92 $\mu_B$) ions, respectively.
These results suggest electron occupancies
of $d^6$ (Fe$^{2+}$) and $d^7$ (Co$^{2+}$), respectively. 
Figure~\ref{band} provides the band structure of 
$\rm Zn_{0.9375}Co_{0.0625}O$ near $E_F$.
Note that the size of circle represents the amount of Co 3$d$
component in the wave function.
It is seen that the rather flat majority spin Co $3d$ states 
are located by $\sim 1.0$ eV below the conduction band 
consisting of mainly Zn 4$s$ states.
The minority spin Co $3d$ states are located near and above $E_F$
manifesting hybridization with the Zn 4$s$ conduction band states.
However, we do not expect that either Fe- or Co-doped ZnO in nature
has a stable metallic ferromagnetic ground state, as obtained above.
The high DOSs at $E_F$ for both Fe- and Co-doped ZnO would drive
the possible structural instability or become reduced substantially
by the Coulomb correlation interaction between TM $3d$ electrons. 
Indeed, LSDA+$U$ band calculation
with $U$=5.0eV for Co $3d$ electrons
yields the insulating ground state for Co-doped ZnO,
distinctly from the LSDA band results\cite{FeZno}.

\section{(F\lowercase{e},C\lowercase{o}) \lowercase{codoped} Z\lowercase{n}O}

Now we have performed the LSDA band calculations for (Fe, Co) codoped ZnO:
$\rm Zn_{0.875}(Fe_{0.5}Co_{0.5})_{0.125}O$. 
We have examined magnetic properties of supercell by varying the
separation between Fe and Co: 3.2499, 5.6055, and 6.4998 $\rm \AA$.
3.2499 $\rm \AA$ corresponds to a nearest Fe-Co 
separation with Fe-O-Co configuration in the a-b plane of the 
orthorhombic supercell, 
while 5.6055 and 6.4998 $\rm \AA$ to farther Fe-Co separations with
Fe-O-Zn-O-Co configurations along the diagonal direction and
in the a-b plane, respectively.
Total energies are nearly the same among three cases:
the shortest 3.2499 $\rm \AA$ case  has the lowest total energy
by only $\sim$ 3 mRy. This result reflects that there will not be 
any noticeable TM clustering effect in the (Fe, Co) codoped ZnO. 
Further we have found that the FM configuration
of Fe and Co spins are found to be slightly more stable than the 
antiferromagnetic (AFM) configuration for all three cases.

Figure ~\ref{zfco} shows the LSDA DOS of $\rm Zn_{1-2x}Fe_{x}Co_{x}O$ 
$(x=0.0625)$ for a Fe-Co separation of 5.6055 $\rm \AA$.
The 3$d$ PLDOS in the codoped case looks as if it is just a simple
sum of Fe and Co PLDOSs of Fig.~\ref{feco}.
There is no indication of charge transfer between Fe and Co,
and so Fe$^{2+}$ and Co$^{2+}$ valence configurations 
are retained in the (Fe, Co) codoped ZnO.
Hence most properties are similar to each Fe- and Co-doped case, such as 
the degree of hybridization, nearly half-metallic 
nature, and the local magnetic moments.
This suggests that the $double$-$exchange$ mechanism\cite{Zener}
will not be effective in $\rm Zn_{1-2x}Fe_{x}Co_{x}O$, 
because the kinetic-energy gain through
the hopping of spin-polarized carriers between Fe and Co ions
does not seem to occur.
Thus to explain the observed FM ground state in $\rm Zn_{1-2x}Fe_{x}Co_{x}O$,
one needs to invoke another exchange mechanism between Fe and Co,
such as the RKKY-type exchange interaction mediated by Zn $4s$
carriers or conduction carriers induced by oxygen vacancies.
However, one cannot rule out the formation of separated
Fe or Co metallic clusters in $\rm Zn_{1-2x}Fe_{x}Co_{x}O$
which would exhibit the ferromagnetism.
Also the possible formation of impurity phases such as
spinel CoFe$_2$O$_4$ is worthwhile to be checked carefully.
These features remain to be resolved more in the experiments.

\begin{figure}[t]
\includegraphics[scale = 0.4]{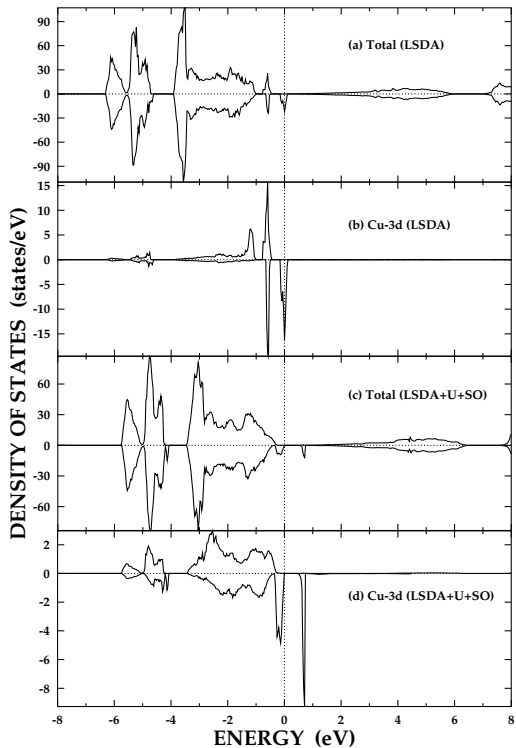}
\caption{The LSDA and LSDA+$U$+SO ($U$=3.0 eV) total and Cu $3d$ 
	PLDOSs for $\rm Zn_{0.9375}Cu_{0.0625}O$.
}
\label{cu}
\end{figure}

\begin{figure}[t]
\includegraphics[scale = 0.4]{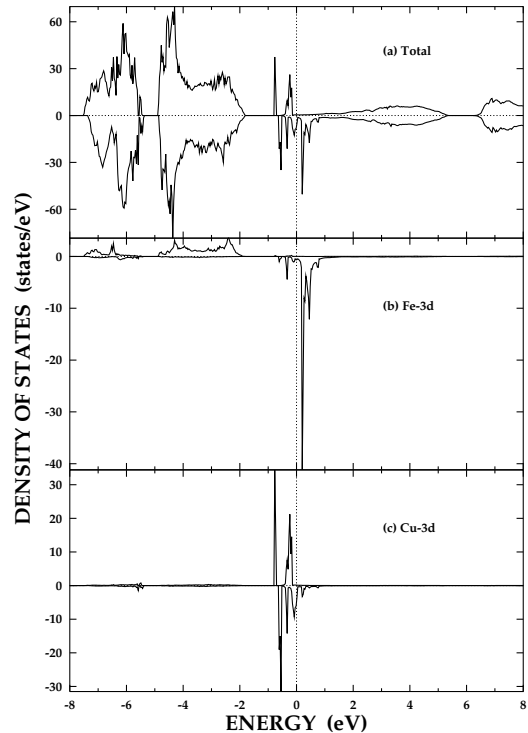}
\caption{The LSDA total and TM $3d$ PLDOSs
         for $\rm Zn_{0.875}Fe_{0.0625}Cu_{0.0625}O$. 
}
\label{zfcuo}
\end{figure}

\section{C\lowercase{u-}\lowercase{doped}  Z\lowercase{n}O}

Before discussing (Fe, Cu) codoped ZnO, we have examined electronic
structure of Cu only  doped ZnO: $\rm Zn_{0.9375}Cu_{0.0625}O$.
Interestingly, as shown in Fig.~\ref{cu},
we have obtained the stable FM and half-metallic 
ground state for Cu-doped ZnO in the LSDA.
The total magnetic moment is 1 $\mu_B$ and the local moment of Cu is 
0.81 $\mu_B$, corresponding to the Cu$^{2+}$ ($d^9$) valence state.
The Cu 3$d$ PLDOS in Fig.~\ref{cu}(b) indicates that Cu has the DOS
of an intermediate spin state. In contrast to Fe- or Co-doped case,
the empty Cu 3$d$ states are located in the gap region 
without the hybridization with the Zn $4s$ conduction band,
and thus Cu 3$d$ states are strongly localized.
We presume that this would be
the reason why the Cu solubility in ZnO is so low: $\sim 1 \%$ at most.

Since the Cu 3$d$ states near $E_F$ correspond to partially occupied
atomic-like $t_{2g}$ states,
Cu ions would have large orbital magnetic moment.
Indeed the LSDA+$U$+SO calculation 
yields the substantial orbital magnetic moment of 1.05$\mu_B$ 
with the insulating electronic structure\cite{Ujcu}.
The large orbital moment arises from occupied minority 
spin $t_{2g}$ states split by the Coulomb correlation and the spin-orbit 
interaction\cite{Mspark}.
The orbital moment is polarized in parallel with the spin 
moment, and so the total magnetic moment amounts to 1.85$\mu_B$/Cu.

In general, an ion at the tetrahedral center with low spin $d^9$ state 
would be Jahn-Teller active. In fact, there was a report 
that the account of the dynamical Jahn-Teller effect is necessary to explain
the observed paramagnetic susceptibility for Cu-doped ZnO\cite{Brumage}.
To examine the Jahn-Teller effect in Cu-doped ZnO, we have considered 
the local distortion of a tetrahedron around Cu by tilting oxygen ions at
the corners. The oxygen ions are tilted by $\sim 13^{\circ}$
toward the $xy$-plane with
retaining Cu-O separations (here $xyz$ coordinates represent the local 
principle axis). The LSDA band calculation for this system
yields a more stable insulating ground state 
with enhanced spin magnetic moment
of 0.99$\mu_B$/Cu. That is, due to the Jahn-Teller distortion,
Cu $3d$ electrons become more localized and the system becomes insulating.
The orbital moment in this case would be quenched due to the
Jahn-Teller effect.

\section{(F\lowercase{e},C\lowercase{u}) 
\lowercase{codoped} Z\lowercase{n}O}

As is done for (Fe, Co) codoped case, we have studied magnetic 
properties of (Fe, Cu) codoped $\rm Zn_{1-2x}Fe_{x}Cu_{x}O$
$(x=0.0625)$ with varying the separation between Fe and Cu: 
3.2499, 5.6055, and 6.4998 \AA. 
For all three cases, the FM configurations of Fe and Cu spins are found 
to be more stable than the AFM configurations, as for the (Fe, Co)
codoped case. In contrast to the (Fe, Co) codoped case, however, 
we have found that the shortest Fe-Cu configuration
becomes much more stable with respect to other two cases by 
$\sim$ 30 mRy.
This total energy difference is very large when comparing with $\sim$ 3 mRy
difference for the (Fe, Co) codoped case.
This result indicates that Fe and Cu ions in (Fe, Cu) codoped ZnO
have a tendency to form the Fe-O-Cu clusters.
For 1 $\%$ Cu-doped $\rm Zn_{0.95}Fe_{0.05}O$ bulk sample,  
Han {\it et al.}\cite{Han} observed rather small saturated magnetic 
moment of 0.75 $\mu_B$ per Fe.
This value corresponds to only about 1/5 of ideal Fe$^{2+}$ 
local moment 4 $\mu_B$.
Incidentally, the value 1/5 is matched with Cu/Fe doping ratio, suggesting
a possibility that the only Fe's forming the Fe-O-Cu clusters would 
give rise to the FM moment, that is, other Fe ions do not produce the
long range FM order. 
As shown below, the FM ground state in (Fe, Cu) codoped ZnO
can be understood based on the enhanced
double-exchange-like interaction through the Fe-O-Cu clustering effect. 

Figure ~\ref{zfcuo} presents the LSDA DOS of 
$\rm Zn_{0.875}Fe_{0.0625}Cu_{0.0625}O$
for the shortest 3.2499 $\rm \AA$ configuration.
Above all, one can notice the strong hybridization between Fe and Cu 
$3d$ states. Further, it is clearly seen in Fig. ~\ref{zfcuo}(c) that 
some new states emerge in the Cu 3$d$ minority spin PLDOS 
between $e_g$ and $t_{2g}$.
The new states have $d_{xy}$-like characters 
directing toward Fe ions\cite{Local}.
Cu $3d$ PLDOS has a reduced exchange splitting, manifesting
the DOS characteristic of the low spin state.
This feature is different from other two configurations with 
larger Fe-Cu separation, which have an intermediate spin state 
as in Cu only doped ZnO.
For all three cases, the hybridization between Cu 
$3d$ and the conduction band exists, 
which is again distinct from Cu only doped ZnO.
Due to this hybridization, the conduction carriers of mostly Zn $4s$ 
states become reduced.
Noteworthy from PLDOSs of Fig.~\ref{zfcuo} is that 
there occurs charge transfer 
from Fe to Cu, and accordingly Fe and Cu are likely to 
have nominal Fe$^{3+}$ ($d^5$) and Cu$^{1+}$ ($d^{10}$) configurations, 
respectively.
As a result, the electron occupancy at Cu site increases,
and so Cu has the reduced spin magnetic moment of 0.51 $\mu_B$
as compared to 0.81 $\mu_B$ in Cu only doped ZnO.
This feature explains the experimental result
of the reduced number of carriers in (Fe, Cu) codoped ZnO 
with respect to that in Fe only doped ZnO \cite{Han}.
The charge transfer from Fe to Cu is expected to disturb the Jahn-Teller
distortion at Cu sites and concomitantly make a system metallic.
Further, it will cause the mixed-valent occupancies 
for Fe (Fe$^{2+}$-Fe$^{3+}$) and Cu (Cu$^{2+}$-Cu$^{1+}$) ions, 
and the consequent double-exchange-like interaction
is expected to induce the ferromagnetism in (Fe, Cu) codoped ZnO.

\section{Conclusion}

We have investigated electronic structures of 
(Fe, Co) and (Fe, Cu) codoped ZnO together with those of
Fe-, Co-, and Cu-doped ZnO.
We have also explored origins of observed ferromagnetism
in (Fe, Co) and (Fe, Cu) codoped ZnO.
The single TM-doped systems
would not have stable metallic ferromagnetic ground states,
due to large Coulomb correlation or other
structural instability effects in Fe- and Co-doped ZnO
and the Jahn-Teller effect in Cu-doped ZnO, respectively.
For (Fe, Co) codoped ZnO, we have found no indication 
of charge transfer between Fe an Co, suggesting that the double-exchange 
mechanism will not be effective for the observed ferromagnetism in
(Fe, Co) codoped ZnO. Therefore one needs to invoke another 
exchange mechanism between Fe and Co, or
the possible formation of impurity phases is to be checked carefully.
In contrast, for (Fe, Cu) codoped ZnO, there is
a tendency to form the Fe-O-Cu clusters and so to give rise to
charge transfer from Fe to Cu ion.
This tendency causes the mixed-valent occupancies between Fe and Cu,
and accordingly the double-exchange-like interaction
is expected to induce the ferromagnetism in (Fe, Cu) codoped ZnO.
The FM and nearly half-metallic ground state is obtained
for (Fe, Cu) codoped ZnO.

\acknowledgments 

This work was supported by the KOSEF through the eSSC at POSTECH
and in part by the KRF (KRF--2002-070-C00038).
Helpful discussions with Y. H. Jeong and S. J. Han are greatly appreciated.


\begin{thebibliography}{99}

   \bibitem{Wolf} S. A. Wolf, D. D. Awschalom, R. A. Buhrman, J. M. Daughton, 
        S. von Molnar, M. L. Roukes, A. Y. Chtchelkanova, and D. M. Treger, 
        Science {\bf 294}, 1488 (2001).
   \bibitem{Furdyna} J. K. Furdyna and J. Kossut, {\it DMSs},
        Semiconductor and Semimetals {\bf 25},
        Academic Press, New York, (1988). 
   \bibitem{Ohno} H. Ohno, A. Shen, F. Matsukura, A. Oiwa, A. Endo, 
        S. Katsumoto, and Y. Iye,
        Appl. Phys. Lett. {\bf 69}, 363 (1996).
   \bibitem{Dietl} T. Dietl, H. Ohno, F. Matsukura, J. Cibert, and D. Ferrand,
	Science {\bf 287}, 1019 (2000).
   \bibitem{Sato} K. Sato and H. K. Yoshida,
	Semicond. Sci. Technol. {\bf 17}, 367 (2002).
   \bibitem{Medvedkin} G. A. Medvedkin, T. Ishibashi, T. Nishi, K. Hayata, 
        Y. Hasegawa, and K. Sato,
	Jpn. J. Appl. Phys. {\bf 39}, L949 (2000).
   \bibitem{Matsumoto} Y. Matsumoto, M. Murakami, T. Shono, T. Hasegawa, 
        T. Fukumura, M. Kawasaki, P. Ahmet, T. Chikyow, S. Koshihara, 
        and H. Koinuma, Science {\bf 291}, 854 (2001).
   \bibitem{Ando} K. Ando, H. Saito, Z. Jin, T. Fukumura, M. Kawasaki,
        Y. Matsumoto, and H. Koinuma,
        Appl. Phys. Lett. {\bf 78}, 2700 (2001).
   \bibitem{SCho} S. Cho, S. Choi, G.-B. Cha, S. C. Hong, 
	Y. Kim, Y.-J. Zhao, A. J. Freeman, J. B. Ketterson,
	B. J. Kim, Y. C. Kim, B.-C. Choi, 
	Phys. Rev. Lett. {\bf 88}, 257203 (2002) 
   \bibitem{Jin} Z. Jin, T. Fukumura, M. Kawasaki, K. Ando, H. Saito, 
        T. Sekiguchi, Y. Z. Yoo, M. Murakami, Y. Matsumoto, T. Hasegawa, 
        and H. Koinuma, Appl. Phys. Lett. {\bf 78}, 3824 (2001).
   \bibitem{Ueda} K. Ueda, H. Tabata, and T. Kawai, 
        Appl. Phys. Lett. {\bf 79}, 988 (2001).
   \bibitem{Cho} Y. M. Cho, W. K. Choo, H. Kim, D. Kim, and Y. Ihm, 
        Appl. Phys. Lett. {\bf 80}, 3358 (2002).
   \bibitem{Han} S. -J. Han, J. W. Song, C.-H. Yang, S. H. Park, J.-H. Park, 
        Y. H. Jeong, and K. W. Rhie, Appl. Phys. Lett. {\bf 81}, 4212 (2002).
   \bibitem{Kwon} S. K. Kwon and B. I. Min, 
	Phys. Rev. Lett. {\bf 84}, 3970 (2000).
   \bibitem{Schroer} P. Schr\"{o}er, P. Kruger, and J. Pollmann,
	Phys. Rev. B {\bf 47}, 6971 (1993).
   \bibitem{Arya} M. Oshikiri and F. Aryasetiawan,
	J. Phys. Soc. Japan, {\bf 69}, 2113 (2000).
   \bibitem{Usuda} M. Usuda, N. Hamada, T. Kotani, and M.van Schilfgaarde, 
	Phys. Rev. B {\bf 66}, 125101 (2002).
   \bibitem{Ruckh} M. Ruckh, D. Schmid, and H. W. Schock,
        J. Appl. Phys. {\bf 76}, 5945 (1994).
   \bibitem{Super} Since the Zn 3$d$ bands in the LSDA are 
	located still deep in energy, the effects of the Zn 3$d$ bands on the 
	magnetic properties of TM doped ZnO will be minor.
   \bibitem{FeZno} For Fe-doped ZnO, the LSDA+$U$ band calculation
	with $U=5.0$ eV does not yield the converged results, 
	probably because of too high DOS at $E_F$. 
	With $U$=9.0eV, we can obtain the converged 
	insulating ground state.
   \bibitem{Zener} C. Zener, Phys. Rev. B {\bf 82}, 403 (1951).
   \bibitem{Ujcu} We employed parameter values of $U$=3.0 eV and $J$=0.87 eV 
	($J$: exchange parameter) for Cu 3$d$ electrons.
   \bibitem{Mspark} M. S. Park, S. K. Kwon, and B. I. Min,
	Phys. Rev. B {\bf 65}, 161201(R) (2002).
   \bibitem{Brumage} W. H. Brumage, C. F. Dorman, and C. R. Quade,
        Phys. Rev. B {\bf 63}, 104411 (2001).
   \bibitem{Local} Here too $xyz$ coordinates 
	represent the local principal axis.


\end{thebibliography}
\end{document}